\documentclass[fleqn,usenatbib]{mnras}

\usepackage{newtxtext,newtxmath}
\usepackage[T1]{fontenc}
\usepackage{ae,aecompl}

\usepackage[dvipdfmx]{graphicx}	% Including figure files
\usepackage{amsmath}	% Advanced maths commands
\usepackage{threeparttable}
\usepackage{nccmath}
\usepackage{multirow}
\usepackage{url}
\usepackage{lineno} % line numbers
\usepackage{gensymb}

\newcommand*\patchAmsMathEnvironmentForLineno[1]{
  \expandafter\let\csname old#1\expandafter\endcsname\csname #1\endcsname
  \expandafter\let\csname oldend#1\expandafter\endcsname\csname end#1\endcsname
  \renewenvironment{#1}
     {\linenomath\csname old#1\endcsname}
     {\csname oldend#1\endcsname\endlinenomath}}
\newcommand*\patchBothAmsMathEnvironmentsForLineno[1]{
  \patchAmsMathEnvironmentForLineno{#1}
  \patchAmsMathEnvironmentForLineno{#1*}}
\AtBeginDocument{
\patchBothAmsMathEnvironmentsForLineno{equation}
\patchBothAmsMathEnvironmentsForLineno{align}
\patchBothAmsMathEnvironmentsForLineno{flalign}
\patchBothAmsMathEnvironmentsForLineno{alignat}
\patchBothAmsMathEnvironmentsForLineno{gather}
\patchBothAmsMathEnvironmentsForLineno{multline}
}

\title[Anisotropic satellite quenching up to $z\sim1$]{Detection of anisotropic satellite quenching in galaxy clusters up to $z\sim1$}

\author[M. Ando, K. Shimasaku and K. Ito]{
Makoto Ando,$^{1}$\thanks{E-mail: mando@astron.s.u-tokyo.ac.jp} Kazuhiro Shimasaku$^{1,2}$ and Kei Ito$^{1}$
\\
$^{1}$Department of Astronomy, Graduate School of Science, The University of Tokyo, 7-3-1 Hongo, Bunkyo-ku, Tokyo 113-0033, Japan\\
$^{2}$Research Center for the Early Universe, The University of Tokyo, 7-3-1 Hongo, Bunkyo-ku, Tokyo 113-0033, Japan\\
}

\date{Accepted XXX. Received YYY; in original form ZZZ}

\pubyear{2022}

\begin{document}
\label{firstpage}
\pagerange{\pageref{firstpage}--\pageref{lastpage}}
\maketitle

%\linenumbers

\begin{abstract}
Satellite galaxies in the cluster environment are more likely to be quenched than galaxies in the general field. Recently, it has been reported that satellite galaxy quenching depends on the orientation relative to their central galaxies: satellites along the major axis of centrals are more likely to be quenched than those along the minor axis. In this paper, we report a detection of such anisotropic quenching up to $z\sim1$ based on a large optically-selected cluster catalogue constructed from the Hyper Suprime-Cam Subaru Strategic Program. We calculate the quiescent satellite galaxy fraction as a function of orientation angle measured from the major axis of central galaxies and find that the quiescent fractions at $0.25<z<1$ are reasonably fitted by sinusoidal functions with amplitudes of a few percent. Anisotropy is clearer in inner regions ($<r_\mathrm{200m}$) of clusters and not significant in cluster outskirts ($>r_\mathrm{200m}$). We also confirm that the observed anisotropy cannot be explained by differences in local galaxy density or stellar mass distribution along the two axes. Quiescent fraction excesses between the two axes suggest that the quenching efficiency contributing to the anisotropy is almost independent of stellar mass, at least down to our stellar mass limit of $M_{*}=1\times10^{10}\,M_{\odot}$. Finally, we argue that the physical origins of the observed anisotropy should have shorter quenching timescales than $\sim1\,\mathrm{Gyr}$, like ram-pressure stripping, because, for anisotropic quenching to be observed, satellites must be quenched before their initial orientation angles are significantly changed.

\end{abstract}


%%%%%%%%%%%%%%%%%%%%%%%%%%%%%%%%%%%%%%%%%%%%%%%%%%

%%%%%%%%%%%%%%%%% BODY OF PAPER %%%%%%%%%%%%%%%%%%

\section{Introduction}
Galaxy clusters are the most massive ($M_\mathrm{h}\gtrsim10^{14}\,M_{\odot}$) virialized systems in the Universe (e.g. \citealp{Kravtsov_Borgani_2012,Overzier2016}). They are also the densest environments in the Universe, hosting thousands of galaxies and hot plasma detected in X-ray, known as the intracluster medium (ICM). In the local Universe, galaxies in clusters typically have elliptical morphologies, quenched star formation activities, and old stellar populations, while those in the general field are rather star-forming galaxies with disc-like morphologies (e.g. \citealp{Dressler1980,Bower1998,Wetzel2012,Chartab2020}). The formation history of such a quenched population in clusters is important to understand galaxy evolution in dense environments.

In general, quenching of galaxy star formation depends on redshift, stellar mass and environment (e.g. \citealp{Dressler1980,Butcher1984,Cooper2007,Kodama2007,Peng2010,Wetzel2012,Darvish2016,Fossati2017,Kawinwanichakij2017,Moutard2018,Chartab2020,Lemaux2022}). In particular, the formation of cluster galaxies is strongly affected by environmental effects which are unique to massive host haloes: gas stripping by ram pressure (e.g. \citealp{Gunn1972}), multiple close encounters with other galaxies (harassment; e.g. \citealp{Moore1996,Moore1998}), and cutoff of cold gas supply (strangulation; e.g. \citealp{larson1980}). Since clusters are connected to larger-scale filamentary structures, some fraction of the quenched galaxies in a cluster may have already been quenched in surrounding filaments by environmental effects there before accretion onto the cluster (pre-processing; \citealp{Fujita2004,DeLucia2012,Donnari2021,Werner2022,Kuchner2022}).

The distribution of satellite galaxies in a halo is known to be anisotropic: satellites are tend to be aligned to the major axis of the central galaxy (e.g. \citealp{Yang2006,Wang2008,Huang2016}) and that of its hosting halo (e.g. \citealp{Paz2011}). These alignments are clearer for satellites which are more luminous, redder and closer to the cluster centres \citep{Yang2006,Azzaro2007,Huang2016,Rodriguez2022}. Since the major axis of central galaxies is also aligned to the larger-scale structure around the clusters \citep{Rodriguez2022}, the segregation of red satellites is possibly related to pre-processing within surrounding filaments.

The anisotropic distribution of red satellites can also be interpreted as the result of environmental effects which work anisotropcally within a halo. Recently, it has been reported that satellite galaxy quenching depends on the orientation relative to their central galaxies: satellites along the major axis of centrals are more likely to be quenched than those along the minor axis \citep{Martin-Navarro2021,Stott2022}. This phenomenon, termed ``anisotropic qunehicng'' or ``angular conformity'', may provide clues to understanding how the cluster environment quench satellites.

Using a spectroscopic group sample at $z<0.08$ from the Sloan Digital Sky Survey, \citet{Martin-Navarro2021} have detected an anisotropy in the quiescent fraction. From a comparison with a hydrodynamical simulation, they have concluded that this anisotropy is caused by the feedback from the active galactic nuclei (AGNs) of central galaxies. Because the minor axis of a central is the direction where AGN feedback removes intra-halo gas more efficiently, the quenching effect due to ram-pressure stripping is weakened in this direction, resulting in the anisotropic quenching.

\citet{Stott2022} has extended this work to $z\sim0.5$ with $13$ massive clusters from the Cluster Lensing And Supernova survey with Hubble (CLASH). He has shown that the rest-frame colour in the vicinity of the $4000\,\text{\AA}$ is redder for galaxies around the major axis of central galaxies than those around the minor axis, which is consistent with \citet{Martin-Navarro2021}'s results if the rest-frame colour is used as an indicator of star formation. He has also proposed another scenario to explain anisotropic quenching in terms of alignment between the major axes of centrals and cluster shapes. If the cluster morphology is elliptical, the distributions of the ICM and galaxies are also elongated, and thus, at a fixed radius, the quenching due to ram-pressure stripping or gravitational interaction becomes more efficient in the direction of elongation. Therefore, an alignment between central galaxies and cluster halos may result in a higher quiescent fraction along the major axis. 

Although these pioneering works have shed light on anisotropic quenching, it is still unclear when and how this anisotropy arose. Since the dominant quenching mechanisms in groups and clusters are likely to be different between the local and distant Universes \citep{Peng2010,Balogh2016,vanderBurg2020,Chartab2020,Reeves2021,Ando2022}, there is a possibility that anisotropic quenching also has redshift evolution. In this sense, whether anisotropic quenching exists at higher redshifts is a vital issue for understanding the physical origins of the anisotropy. To tackle this, we need a cluster sample over a wide redshift range sufficient for statistical discussion.

A wide and deep optical imaging survey, named the Hyper Suprime-Cam (HSC; \citealp{Furusawa2018,Miyazaki2018,Komiyama2018}) Subaru Strategic Program (HSC-SSP; \citealp{Aihara2018a,Aihara2022}), provides an ideal data set for this purpose. The HSC-SSP has three observation layers (Wide, Deep and UltraDeep) with different survey areas and depths. The Wide layer, which has the largest area coverage among them, reaches $\sim 1400\,\deg^{2}$ with deep photometry ($i\sim26$; \citealp{Aihara2022}).

In this paper, we report the detection of anisotropic quenching up to $z\sim1$ with a large cluster sample constructed from the HSC-SSP data. We calculate the quiescent satellite galaxy fraction as a function of the orientation angle measured from the major axis of centrals and evaluate the detected signals by model fitting. We also examine the dependence of the detected anisotropy on cluster-centric-radius, local density, and stellar mass.

The structure of this paper is as follows. In \S~2, we describe the data and sample selection of clusters and galaxies from the parent catalogues. In \S~3, we explain how to calculate the position angle (PA) of centrals and the quiescent fraction. In \S~4, we detect anisotropy in quiescent fraction. In \S~5, we constraint the physical origins of the detected anisotropic quenching. \S~6 is devoted to a summary and conclusions.
 
Throughout this paper, we assume a flat $\mathrm{\Lambda}$CDM cosmology with $(\Omega_\mathrm{m},\, \Omega_\mathrm{\Lambda},\, h)=(0.3,\, 0.7,\, 0.7)$ and a \citet{Chabrier2003} initial mass function. We use AB magnitudes \citep{Oke1983} and physical distances.

\section{Data and sample selection}
We use data from the S20A and S21A internal Data Release of the HSC-SSP (\citealp{Aihara2018a,Aihara2022})\footnote{As of this writing, only the S20A data are publicly available (\url{https://hsc-release.mtk.nao.ac.jp/doc/}). Newer data including the S21A data will be open to the community in the future.}. The HSC-SSP is a wide-field optical imaging survey with five broad bands (\textit{grizy}) and four narrowbands \citep{Kawanomoto2018}. Raw images are processed with \texttt{hscPipe 8} \citep{Bosch2019}, which is based on the pipeline of the Large Synoptic Survey Telescope (LSST; \citealp{Juric2017,Ivezic2019}).

The HSC-SSP has three observation layers (Wide, Deep and UltraDeep) with different survey areas and depths. In this study, we focus on the Wide layer, which has the largest area coverage ($\sim 1400\,\deg^{2}$).

\subsection{Galaxy sample}
We use an S21A photo-\textit{z} catalogue based on a spectral energy distribution (SED) fitting code with Bayesian priors on physical properties of galaxies, called \textsc{MIZUKI} \citep{Tanaka2015,Tanaka2018,Nishizawa2020}. We first select galaxies in the ``full-colour'' region, where all five-band data exist. We require at least two exposures for the \textit{g}, \textit{r}, and \textit{i} bands and four exposures for the \textit{y} and \textit{z} bands.\footnote{$\texttt{inputcount\_value}\geq2$ and $\geq4$, respectively, where \texttt{inputcount\_value} is the number of exposures contributing to the co-add image of each galaxy (cf. \citealp{Oguri2018}).} Then, we exclude all objects located in a low-quality region\footnote{The corresponding flags in the catalogues are \texttt{pixelflags\_offimage}, \texttt{pixelflags\_edge}, and \texttt{pixelflags\_bad}.} or contaminated by unreliable pixels, cosmic rays, or bright stars\footnote{The corresponding flags are \texttt{pixelflags\_interpolatedcenter}, \texttt{pixelflags\_saturatedcenter}, \texttt{pixelflags\_crcenter}, \texttt{mask\_brightstar\_halo}, \texttt{mask\_brightstar\_blooming}, \texttt{mask\_brightstar\_ghost15}, and \texttt{mask\_brightstar\_channel\_stop} (only for the y band).} \citep{Bosch2018,Aihara2022}. The effective survey area after these maskings is $\sim 800\,\deg^{2}$.

From the catalogue, we exclude objects with \texttt{cmodel\_flag} to ensure secure CModel magnitudes \citep{Abazajian2004,Bosch2018}. We apply a somewhat strict magnitude cut in the \textit{i} band, $i\leq25.0$, while the $5\sigma$ detection limit for point sources in the \textit{i} band is $i=26.2$. In addition, we also apply a more relaxed magnitude cut for the \textit{z} and \textit{y} bands: $z\leq25.4$, $y\leq24.6$, which corresponds to the $3\sigma$ detection limits. We do not apply a magnitude cut for the \textit{g} and \textit{r} bands to include red (quiescent) galaxies at higher redshifts in our analysis. To ensure secure photo-\textit{z}'s and physical parameters, we only use galaxies with $\texttt{risk}<0.1$, where \texttt{risk} roughly indicates the probability of the photo-\textit{z} being an outlier (i.e. $[z_\mathrm{phot}-z]/[1+z]\gtrsim0.15$, see equations~(12) and (13) in \citealp{Tanaka2018} for more detail), and this criterion drops $\sim30$ per cent of galaxies.

From the catalogue, we extract photo-\textit{z}'s, stellar masses, and star formation rates (SFRs) and select galaxies at $0<z<1.4$. We use specific SFR (sSFR; $\mathrm{SFR}/M_{*}$) to classify galaxies. We define galaxies with $\mathrm{sSFR}<10^{-11}\,\mathrm{yr^{-1}}$ as quiescent galaxies (QGs) and the others as star-forming galaxies (SFGs).

\begin{figure}
	\includegraphics[width=\columnwidth]{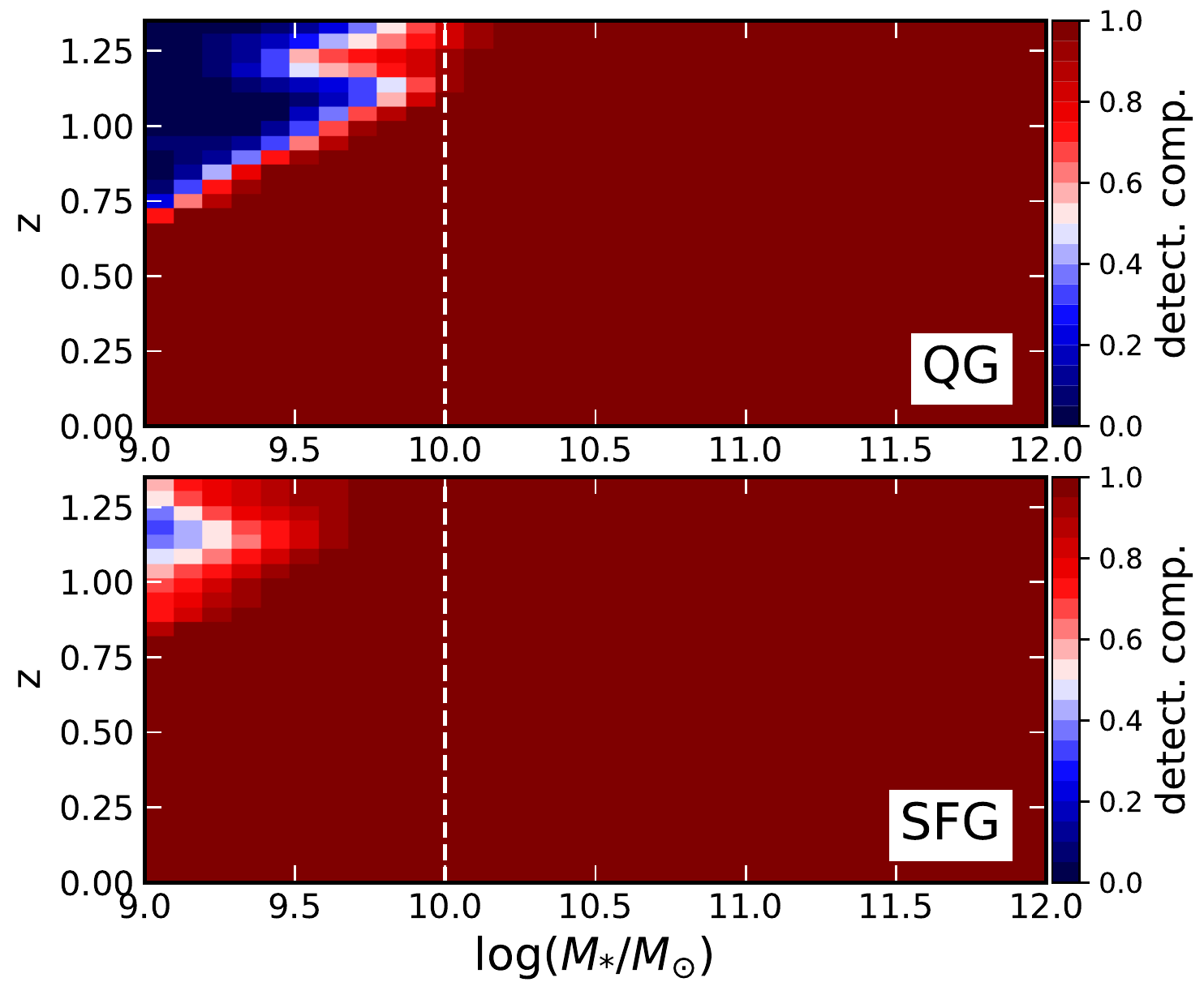}
    \caption{The detection completeness for QGs (top) and SFGs (bottom) as a function of redshift and stellar mass. Both QGs and SFGs are complete above $\log(M_{*}/(M_{\odot})=10$.}
    \label{fig:mass-comp}
\end{figure}

We then examine stellar mass completeness both for QGs and SFGs as a function of stellar mass and redshift by an empirical method (e.g. \citealp{Pozzetti2010, Davidzon2013, Ilbert2013, Weaver2022, Ando2022}). For each galaxy, we calculate rescaled stellar mass $M_\mathrm{*,res}$ defined as:
\begin{equation}
    M_\mathrm{*,res}=\log(M_{*})-0.4(i_\mathrm{lim}-i)
\end{equation}
where $i$ and $i_\mathrm{lim}$ are the observed and limiting \textit{i}-band magnitudes, respectively. Then, the detection completeness at a given redshift $z$ and stellar mass $M$ is estimated as:
\begin{equation}
    \mathrm{detect.\ comp.}(z, M)=\frac{N(z,M_\mathrm{*,res}<M)}{N(z)},
\end{equation}
where $N(z,M_\mathrm{*,res}<M)$ is the number of galaxies with $M_\mathrm{*,res}<M$, and $N(z)$ is the total number of galaxies at this redshift. Fig.~\ref{fig:mass-comp} shows stellar mass completeness for QGs (top) and SFGs (bottom). For both categories, galaxies more massive than $\log(M_{*}/M_{\odot})=10$ are almost complete across all redshifts. Therefore, we limit our sample to galaxies more massive than $\log(M_{*}/M_{\odot})>10$.

\subsection{Cluster sample and membership}
We use the latest version of the CAMIRA (Cluster-finding Algorithm based on Multi-band Identification of Red-sequence gAlaxies) cluster catalogue with bright star masks constructed by applying the CAMIRA algorithm \citep{Oguri2014} to the HSC-SSP S20A data \citep{Oguri2018}\footnote{\url{https://github.com/oguri/cluster_catalogs/tree/main/hsc_s20a_camira}}. CAMIRA clusters are identified mainly at $0.1<z<1.1$ as peaks in the three-dimensional richness map of red-sequence galaxies. CAMIRA clusters have been validated using spectroscopically confirmed galaxies, X-ray observation, and mock galaxies.

In the CAMIRA algorithm, the centre of a cluster is defined as the position of the central galaxy candidate of that cluster. To determine central galaxies in our galaxy sample, we first cross-match our galaxy sample and the CAMIRA cluster catalogue within $1\arcsec$ separation. No central galaxies are found for about ten per cent of the CAMIRA clusters mainly due to stricter bright star masks for our galaxy sample than those used for CAMIRA cluster identification. We exclude these CAMIRA clusters without central galaxies in our analysis.

Using the mass-richness relations of CAMIRA clusters presented in \citet{Murata2019}, we estimate $M_\mathrm{200m}$ and $r_\mathrm{200m}$, indicators of virial mass and virial radius, respectively. Here, $M_\mathrm{200m}$ is the total mass enclosed by the radius, $r_\mathrm{200m}$, whose inner mass density is $200$ times the mean density of the universe. To avoid the mixture of galaxies belonging to different clusters, we exclude cluster pairs whose projected and redshift separations are less than the sum of each cluster's $r_\mathrm{200m}$ and $0.05\times(1+z)$, respectively.

As described in \S~\ref{sec:pa}, we estimate the PA of the central galaxy of each CAMIRA cluster by two-dimensional light profile fitting. We do not use clusters with bad PA estimates. We finalise our cluster sample by removing clusters with the total coverage of masked regions within $r_\mathrm{200m}$ exceeding 40 per cent.

\begin{figure}
	\includegraphics[width=\columnwidth]{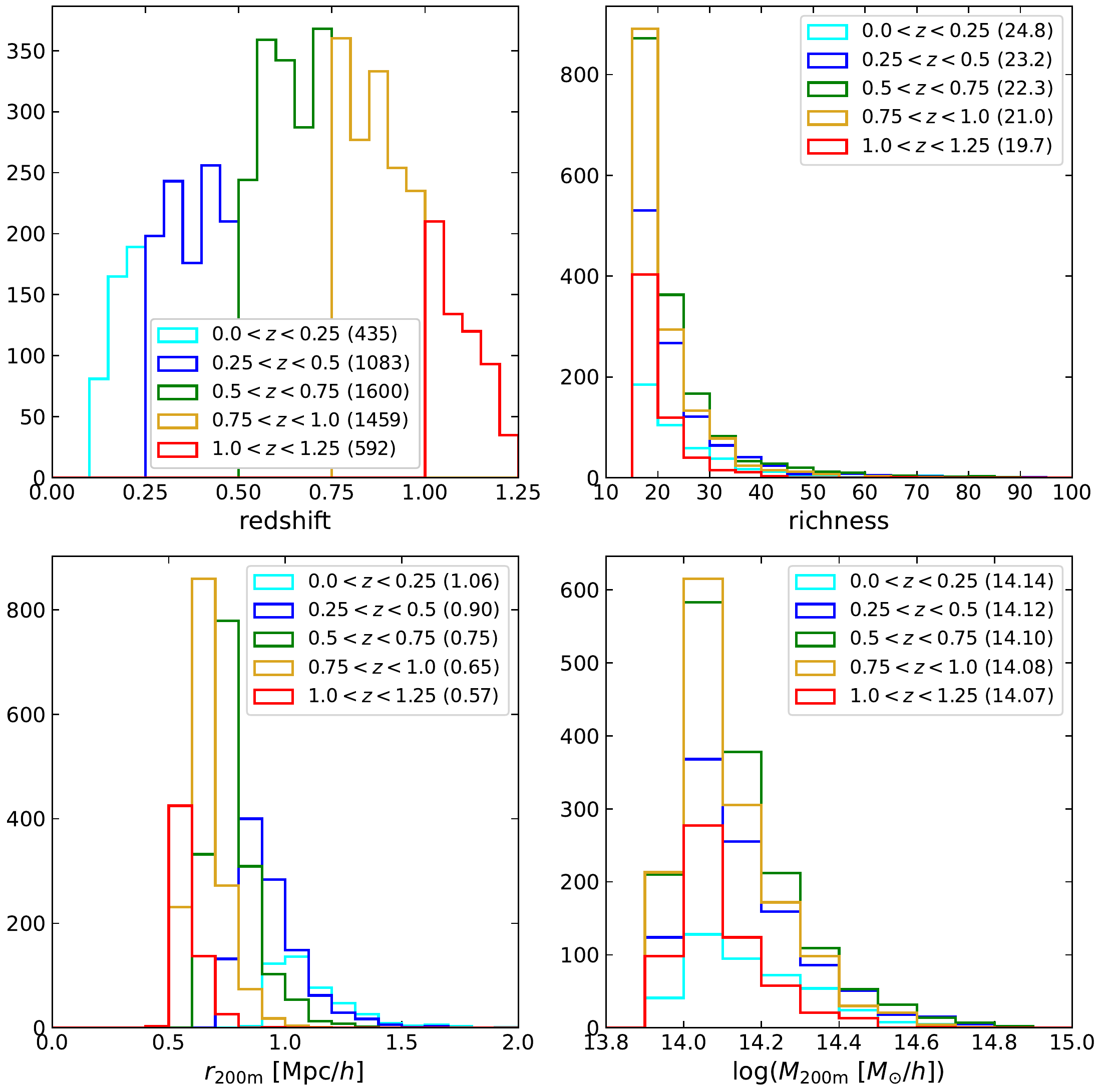}
    \caption{The distributions of redshift (top left), richness (top right), richness inferred $r_\mathrm{200m}$ (bottom left) and $M_\mathrm{200m}$ (bottom right) of our cluster sample colour-coded by redshift bin. The numbers in parentheses in each legend show the number of each redshift subsamples (top left) and the median values of each quantities (other panels).}
    \label{fig:cls_prop}
\end{figure}

We divide the cluster sample into five redshift bins between $0<z<1.25$ with an interval of $\Delta z=0.25$. Each bin contains $\sim 400\text{--}1600$ clusters. The redshift, richness, $M_\mathrm{200m}$ and $r_\mathrm{200m}$ distributions of our cluster sample are shown in Fig.~\ref{fig:cls_prop}.

To determine the member galaxies of a given cluster, we select all galaxies in the cylindrical region around the cluster centre with a radius of $\Delta r\leq 2\times r_\mathrm{200m}$ and a sight-line length of $|\Delta z|\leq0.05\times(1+z_\mathrm{cls})$, where $z_\mathrm{cls}$ is the redshift of the cluster. The sight-line criterion is determined by the typical uncertainty in photo-\textit{z} estimates \citep{Tanaka2018,Nishizawa2020}. If a certain galaxy is included in more than one cluster, that galaxy is assigned to the cluster with the smallest projected separation.

\section{Analyses}

\subsection{Measuring the position angles of central galaxies}
\label{sec:pa}
We measure the PAs of the central galaxies of CAMIRA clusters from two-dimensional light profile fitting using \textsc{GALFIT} \citep{Peng2002_gf,Peng2010_gf}. Since the \textit{i}-band images have the best seeing ($\sim0.6\arcsec$) among the five bands, we fit the \textit{i}-band images of central galaxies ($17\arcsec \times 17\arcsec$) with a single S{\'e}rsic profile model \citep{Sersic1963} after masking other galaxies\footnote{We make masks based on segmentation maps derived from the python package \textsc{astropy.photutils} (\url{https://photutils.readthedocs.io/en/stable/}).}. We leave all five parameters free during fitting: total magnitude, effective radius, S{\'e}rsic index, axis ratio and PA. To ensure secure PA values, we only use centrals whose reduced chi-squares ($\chi_\mathrm{\nu}^{2}$) and uncertainties in PA estimates are less than $3$ and $10\degree$, respectively. This cut drops $\sim7$ per cent of all the centrals, of which $\sim3$ per cent is due to PA uncertainties. We note that $90$ per cent of all the centrals have uncertainties less than $5\degree$.

\subsection{Quiescent fraction}
To calculate the average quiescent fraction among clusters as a function of the position in clusters, we first calculate the orientation angle of satellite galaxies measured from the major axis of central galaxies. Then we normalise the cluster-centric-radius of satellites by the $r_\mathrm{200m}$ of each cluster. Then, we ``stack'' clusters in a given redshift bin as a single pseudo-cluster by aligning the major axes of the central galaxies of individual clusters, i.e. we assign a new position to each satellite as:
\begin{flalign}
\label{eq:newpos_x}
        x=(r/r_\mathrm{200m})\times \cos(\theta),\\
\label{eq:newpos_y}
        y=(r/r_\mathrm{200m})\times \sin(\theta),
\end{flalign}
where $r$ and $\theta$ are the cluster-centric-radius and the orientation angle. In principle, all quantities related to orientation angle should be symmetrical in every $90\degree$.

Since we use a photo-\textit{z} catalogue, these stacked clusters contain fore-/back-ground contaminants. We subtract them in a statistical manner. Considering masked regions, for individual clusters, we first calculate the volumes occupied by cylinders used for determining cluster membership. Then, we estimate the expected numbers of fore-/back-ground contaminants by multiplying those volumes and the galaxy number density calculated from the whole survey volume. Finally, we subtract the sum of these expected numbers of contaminants from the stacked number counts of cluster galaxies. When we consider angular bins, we rescale the amount of subtraction according to the bin size.

To take the uncertainties from field subtraction and cluster-to-cluster variation into account, we estimate the quiescent satellite galaxy fraction and its uncertainty by the Jackknife technique. For a given redshift bin containing $N$ clusters, we first calculate the quiescent fraction using the stacked cluster as described above, $f_\mathrm{q,N}$, by dividing the stacked number of QGs by that of all galaxies after field subtraction. Then, we calculate the quiescent fraction again, but this time we subtract the contribution of the $i$-th cluster from the stacked one, $f_\mathrm{q,-i}$. We repeat this calculation for all clusters and obtain $N$ values of $f_\mathrm{q,-i}$. The quiescent fraction ($f_\mathrm{q}$) and its uncertainty ($\delta f_\mathrm{q}$) are estimated as:
\begin{flalign}
\label{eq:jackknife}
        f_\mathrm{q} &=\frac{1}{N}\sum_{i}^{N}\left[N f_\mathrm{q,N}-(N-1)f_\mathrm{q,-i}\right],  \\[6pt]
        \delta f_\mathrm{q} &=\sqrt{\frac{N-1}{N} \sum_{i}^{N} \left(f_\mathrm{q,-i}-f_\mathrm{q,N}\right)^{2}}.
\end{flalign}

\section{Results}
\label{sec:result}
\subsection{Angular dependence}
\label{sec:angular}
\begin{figure*}
	\includegraphics[width=2\columnwidth]{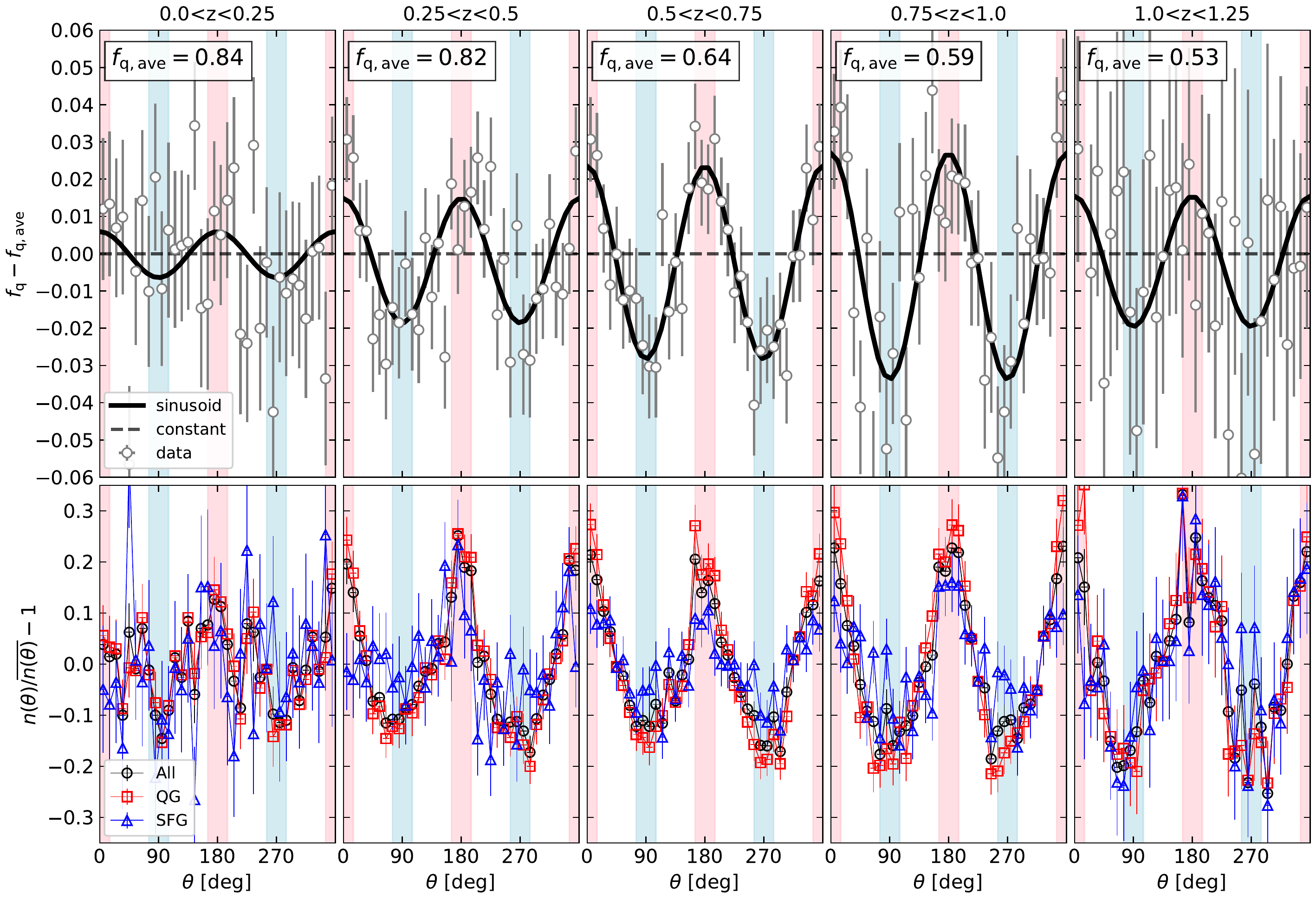}
    \caption{\textit{Top}: The quiescent fraction as a function of the orientation angle measured from the major axis of central galaxies. Zero points are shifted to the average values (shown in the top-left corners of each panel) for clarity. Each column shows different redshifts. Black solid lines and dashed lines are the best-fit models of sinusoids and constant values, respectively. Red and blue shades mean the major axis ($\theta\sim0\degree,\,180\degree,\,360\degree$) and minor axis ($\theta\sim90\degree,\,270\degree$) with $30\degree$ interval. \textit{Bottom}: The angular over density $\delta^{j} (\theta)=n^{j}(\theta)/\overline{n^{j}(\theta)}-1$ for three star-formation categories ($j=\{\mathrm{All|QG|SFG}\}$). A positive (negative) value indicates that galaxies tend to prefer (avoid) the given direction, respectively.}
    \label{fig:angular}
\end{figure*}

In the top panels of Fig.~\ref{fig:angular}, we show the quiescent fractions within $r_\mathrm{200m}$ as a function of orientation angle with $\Delta\theta=10\degree$ bins. The quiescent fractions are offset by the average values just for clarity. For all three redshift bins over $0.25<z<1$, the quiescent fractions fluctuate periodically and are clearly larger around the major axis ($\theta=0\degree$, $180\degree$, $360\degree$) than around the minor axis ($\theta=90\degree$, $270\degree$). Such periodic fluctuations are unclear at $z<0.25$ and $z>1$. As \citet{Martin-Navarro2021} have done, we fit these fluctuations with the sinusoidal function:
\begin{equation}
    f_\mathrm{q}=A_\mathrm{q} \cos(2\theta)+f_\mathrm{q,0},
\end{equation}
where $A_\mathrm{q}$ and $f_\mathrm{q,0}$ are the amplitude of modulation and the mean quiescent fraction, respectively. For comparison, we also fit the quiescent fraction with a constant value, $f_\mathrm{q,const}$. The fitting results are summarised in Table~\ref{tab:fit_result}, and the best-fit models are shown in the top panels of Fig.~\ref{fig:angular}.

At $0.25<z<1$, the amplitudes of modulation are very significant ($>5\sigma$ level). Moreover, sinusoids fit the data ($\chi_{\nu}^{2}\sim1$) much better than constant values ($\chi_{\nu}^{2}\gtrsim2$). This is clear evidence of anisotropic quenching in clusters in a wide redshift range up to $z\sim1$. On the contrary, we do not detect anisotropy at $z<0.25$ with a less significant amplitude ($\sim 1\sigma$ level) and almost the same $\chi_{\nu}^{2}$ ($\sim1$) between the two models. At $z>1$, although the amplitude suggests marginal detection ($\sim 2\sigma$ level), the $\chi_{\nu}^{2}$ value does not clearly prefer the sinusoidal model.

Despite our non-detection of anisotropy at $z<0.25$, \citet{Martin-Navarro2021} have detected significant anisotropy at $z<0.08$ based on a much larger group and cluster sample ($\sim30000$) with spectroscopic redshifts. Since our sample size at $z<0.25$ is relatively small ($\sim400$), non-detection of anisotropy may be due to small statistics rather than due to non-existence of anisotropy. For clusters at $z>1$, we cannot conclude whether the anisotropy exists.

An anisotropy in quiescent fraction means that the spatial distributions in terms of orientation angle are different between QGs and SFGs. To quantify such an angular distribution, we define angular overdensity as:
\begin{equation}
\label{eq:ang_od}
    \delta^{j} (\theta)=\frac{n^{j}(\theta)}{\overline{n^{j}}}-1,
\end{equation}
where $n^{j}(\theta)$ and $\overline{n^{j}}$ are the number of galaxies at a given orientation angle $\theta$ and a given star formation category $j$ (i.e. QG, SFG, all galaxies) and its mean, respectively. We show $\delta^{j} (\theta)$ in the bottom panels of Fig.~\ref{fig:angular}. In any redshift bin, the distribution of QGs is slightly anisotropic, with an excess (deficit) toward the major (minor) axis of centrals at amplitudes of $10\text{--}20$ per cent. Except for the lowest redshift bin, SFGs also have a similar anisotropic distribution but with lower amplitudes, thus causing the observed periodic anisotropy in quiescent fraction. These anisotropic galaxy distributions have also been reported in low-redshift clusters (e.g. \citealp{Yang2006,Huang2016}).

\begin{table*}
\centering
    \caption{The best-fit parameters for quiescent fraction as a function of orientation.}
    \label{tab:fit_result}
    %%% make table %%%
	\begin{tabular}{cc|ccc|cc}
		\hline
		$z_\mathrm{cls}$ & $N_\mathrm{cls}$ & $A_\mathrm{q}$ & $f_\mathrm{q,0}$ & $\chi_{\nu}^2$ (sinusoid) & $f_\mathrm{q,const}$ & $\chi_{\nu}^2$ (constant) \\
		 & \# & $10^{-2}$ & $10^{-2}$ & & $10^{-2}$ &  \\
		\hline
        $[0.10,0.25)$ & $435$ &
        $0.62{\pm0.48}$ & $83.55{\pm0.34}$ & $0.871$ & $83.57{\pm0.34}$ & $0.893$  \\  [3pt]

        $[0.25,0.50)$ & $1083$ &
        $1.67{\pm0.32}$ & $81.69{\pm0.23}$ & $1.049$ & $81.88{\pm0.23}$ & $1.808$ \\  [3pt]
        
        $[0.50,0.75)$ & $1600$ &
        $2.59{\pm0.30}$ & $63.67{\pm0.21}$ & $0.630$ & $63.91{\pm0.21}$ & $2.746$\\  [3pt]
        
        $[0.75,1.00)$ & $1459$ &
        $3.03{\pm0.41}$ & $59.50{\pm0.29}$ & $1.134$ & $59.84{\pm0.29}$ & $2.662$\\  [3pt]
        
        $[1.00,1.25)$ & $592$ &
        $1.75{\pm0.82}$ & $52.95{\pm0.59}$ & $0.567$ & $53.16{\pm0.58}$ & $0.682$\\  [3pt]
        \hline
	\end{tabular}
\end{table*}

\subsection{Radial dependence}
\label{sec:radial}
As shown in \S~\ref{sec:angular}, the quiescent fraction in clusters clearly depends on orientation angle. To see on what scales anisotropic quenching exists, we calculate the quiescent fraction along the major and minor axes as a function of the projected distance from the cluster centre. Here, we exclude clusters with the total coverage of masked regions at $1<r/r_\mathrm{200m}<2$ exceeding 40 per cent and only use galaxies within $\pm 15\degree$ from the major or minor axis direction.

\begin{figure*}
	\includegraphics[width=2\columnwidth]{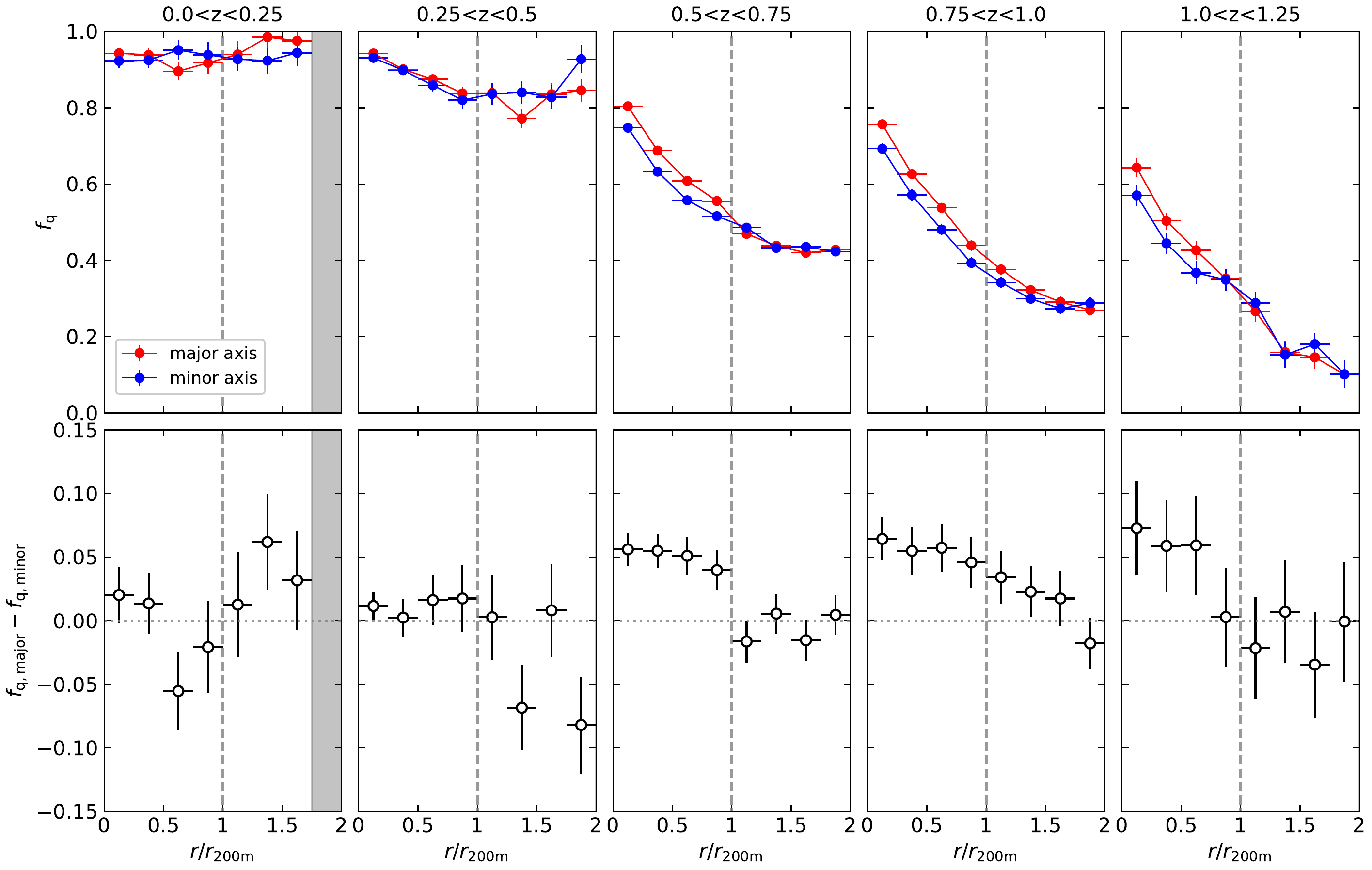}
    \caption{\textit{Top}: The quiescent fraction as a function of cluster-centric-radius. Red and blue symbols indicate the major and minor axis directions with an opening angle of $30\degree$, respectively. Each column shows different redshifts. Vertical dashed lines are the positions of $r_\mathrm{200m}$. Data points with negative galaxy number counts due to field subtraction are not shown, and that region is indicated by a grey shade. \textit{Bottom}: The difference in quiescent fraction between the major and minor axes.}
    \label{fig:radial}
\end{figure*}

The radial profiles of quiescent fraction (top) and their differences between the two axes (bottom) are shown in Fig.~\ref{fig:radial}. Except for the lowest redshift bin, quiescent fraction depends on cluster-centric-radius, decreasing toward the cluster outskirt. At any radius within $r_\mathrm{200m}$, the quiescent fraction along the major axis exceeds that along the minor axis at these redshifts. On the contrary, in the cluster outskirt ($1<r/r_\mathrm{200m}<2$), the quiescent fractions along the two axes are almost the same within uncertainties. These results suggest that the physical mechanisms causing anisotropic quenching mainly work within a cluster halo rather than within the large-scale structure outside of the cluster as ``pre-processing''. We note that the anisotropy signal within $r_\mathrm{200m}$ seems significant even at $1<z<1.25$ although the observed anisotropy shown in \S~\ref{sec:angular} is very tentative at this redshift. This suggests that anisotropy may exist if one focuses only near the major and minor axes.

\citet{Martin-Navarro2021} have reported that the amplitudes of anisotropic quenching are larger for inner regions (within $0.5$ times of the virial radii) compared to outer regions. This trend is not clearly seen in our sample, but is consistent with our results that the signals of anisotropic quenching becomes insignificant outside of $r_\mathrm{200m}$.

One caveat is that there is a possibility that the anisotropic quenching in cluster outskirts indeed exists but is hidden by field contaminants, which have an isotropic spatial distribution against the central's major axis, because the galaxy density in cluster outskirts is much lower than that in the innermost region. This is not a problem if we consider only the regions within $r_\mathrm{200m}$, where cluster galaxies are dominant against contaminants. Indeed, cluster member galaxies occupy $\sim 65\text{--}95$ per cent of all galaxies within $r_\mathrm{200m}$ with larger dominance for lower redshifts. In any case, we need to spectroscopically confirm member galaxies to avoid this possible systematics.

\section{Discussion}
\subsection{Anisotropic quenching versus local density}
\label{sec:local_density}
In general, galaxies located in denser environments have higher quiescent fractions (e.g. \citealp{Peng2010,Darvish2016}). As shown in the bottom panels of Fig.~\ref{fig:angular}, satellite galaxies are preferentially distributed along the major axis of centrals, suggesting that the local galaxy number density is higher along this direction within $r_\mathrm{200m}$. Therefore, there is a possibility that the observed modulation in quiescent fraction is merely a reflection of different local densities between the two directions (cf. \citealp{Stott2022}). Motivated by this idea, we test whether the observed anisotropy in quiescent fraction can be fully explained by the difference in the local density along the two axes.

For each galaxy within $r_\mathrm{200m}$ of a given cluster, we estimate the local galaxy number density with the two-dimensional Gaussian kernel \citep{Ito2021}:
\begin{equation}
    \label{eq:kd}
    \rho_{i}=\sum_{j}\frac{1}{2\pi a^{2}}\exp \left( -\frac{d_{i,j}}{2a^{2}} \right),
\end{equation}
where $a$ is the smoothing scale of the Gaussian kernel and $d_{i,j}$ is the projected distance of galaxy \textit{j} from the galaxy in question. In this calculation, we take summation for all galaxies in the cluster region within $0.05\times(1+z_{i})$ from the galaxy in question at $z_{i}$. Selecting the smoothing scale $a$ is to determine the scale of interest. To balance between small statistics due to a small kernel size and a poor spatial resolution due to a large smoothing scale, we adopt the mean galaxy separation within clusters at each redshift bin calculated as follows. For each redshift bin, we first calculate the average galaxy separation in individual clusters as $l_{i}=\sqrt{{N_{i}}/{\pi r_{\mathrm{200m},i}^{2}}}$, where $N_{i}$ and $r_{\mathrm{200m},i}$ are the richness and $r_\mathrm{200m}$ of cluster \textit{i}. Then we take the average of $l_{i}$ and adopt it as the smoothing scale of this redshift bin. The estimated average separations range over $0.235\text{--}0.405\,\mathrm{Mpc}/h$ for the five redshift bins.\footnote{We note that this is not the actual mean separation but rough estimation of it because the richness of CAMIRA clusters is estimated only from red-sequence galaxies, and more practically, our cluster membership and that of the CAMIRA catalogue are not perfectly matched. However, we confirm that our results are qualitatively not changed if we adopt fixed scales of $0.1$, $0.3$ or $0.5\, \mathrm{Mpc}/h$.}

Then, we define the overdensity ($\delta_{i}$) around galaxy \textit{i}:
\begin{equation}
\label{eq:overdesnity}
    \delta_{i} =\frac{\rho_{i}}{\overline{\rho}}-1,
\end{equation}
where $\overline{\rho}$ is the mean value of the kernel density calculated over galaxies in all clusters at a given redshift bin. Basically, galaxies at $\delta\gtrsim 2$ are mainly located at inner regions of clusters ($\lesssim 0.2\times r_\mathrm{200m}$), where galaxies along the two axes can be mixed within the smoothing scale. We note that we do not perform field subtraction because we cannot know how many contaminants are in each density bin.

\begin{figure*}
	\includegraphics[width=2\columnwidth]{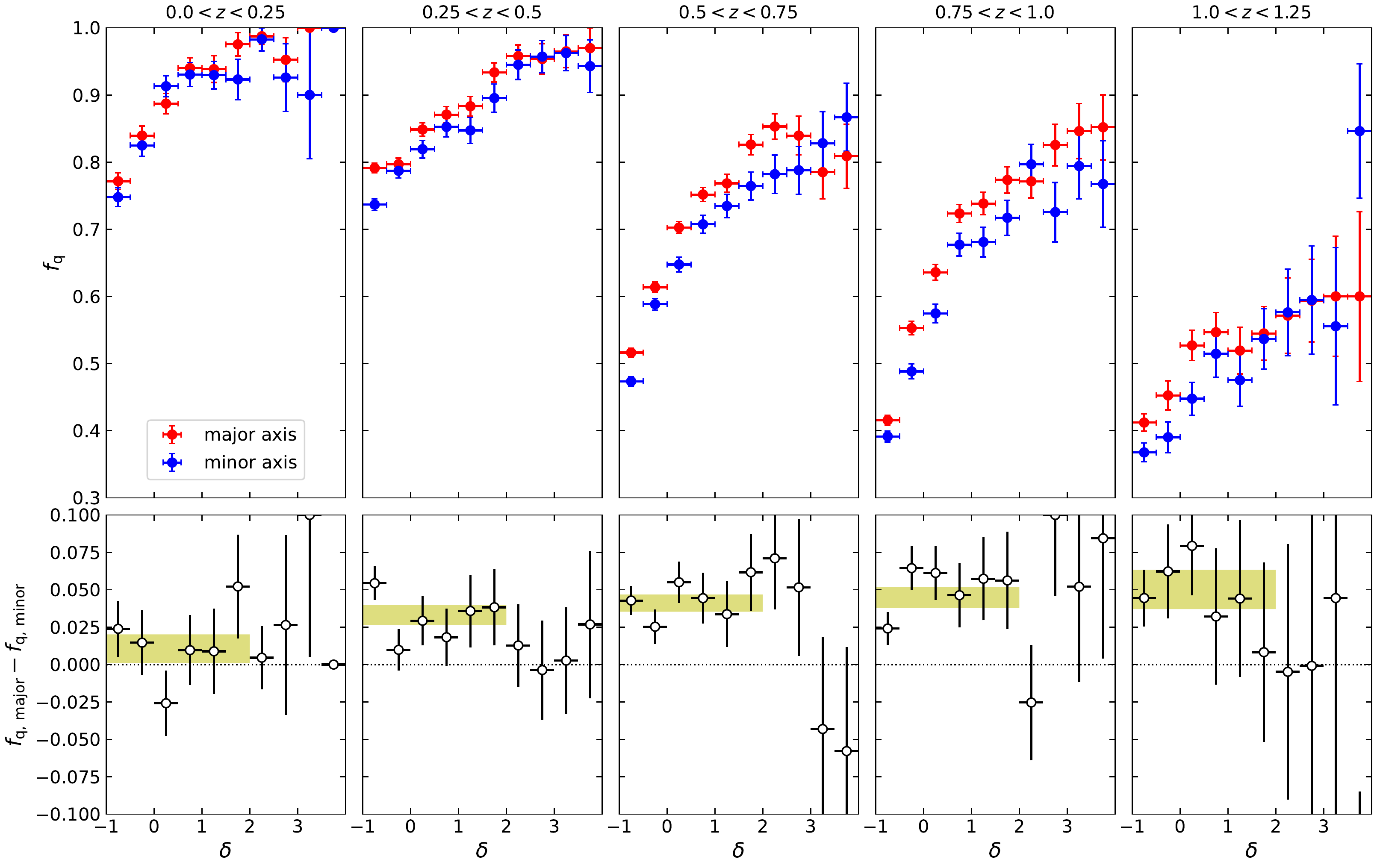}
    \caption{\textit{Top}: The quiescent fraction as a function of the local overdensity  $\delta_{i}$ measured within individual clusters as equations~\eqref{eq:kd} and \eqref{eq:overdesnity}. Red and blue symbols indicate the major and minor axis directions with an opening angle of $30\degree$, respectively. Each column shows different redshifts. \textit{Bottom}: The Difference in quiescent fraction between the two axes at each overdensity bin. Yellow shades indicate the average values with their standard errors over $-1<\sigma<2$, where satellites along the major and minor axes are not clearly mixed.}
    \label{fig:overdensity}
\end{figure*}

We plot the quiescent fraction as a function of the overdensity in the top panels of Fig.~\ref{fig:overdensity}. For both axes, the quiescent fraction increases with the local density, which is interpreted as a reflection of the relation between the star formation and the local density (e.g. \citealp{Darvish2016,Kawinwanichakij2017}). If we compare the two axes at a fixed overdensity, the quiescent fraction is slightly but significantly higher along the major axis (Fig.~\ref{fig:overdensity}, bottom). Moreover, the average values over $-1<\delta<2$, shown by yellow shades, significantly exceed zero except for the lowest redshift bin.

\begin{figure*}
	\includegraphics[width=2\columnwidth]{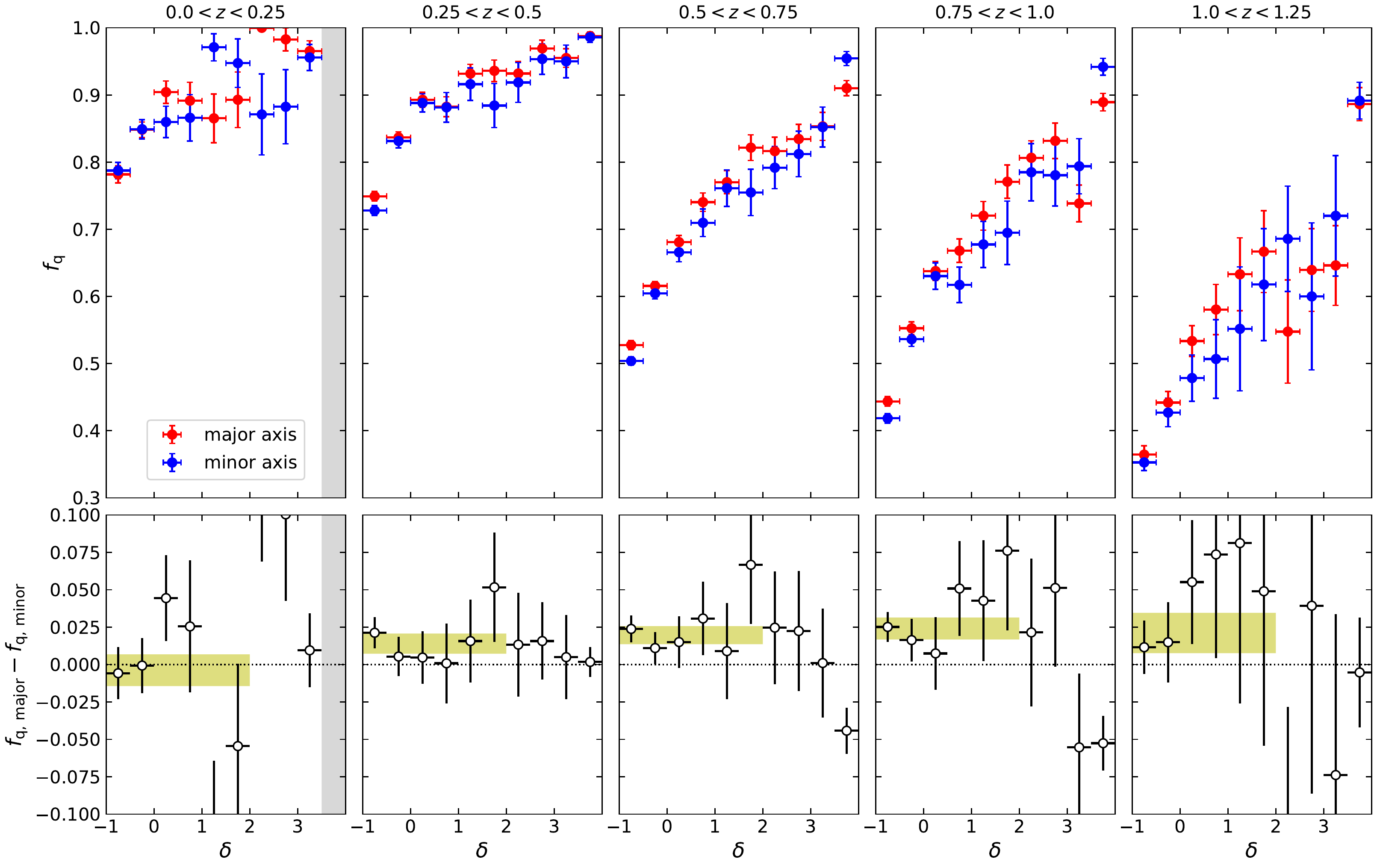}
    \caption{The same as Fig.~\ref{fig:overdensity}, but the local density is measured within the stacked clusters. The $\sigma$ ranges with no data points are indicated by grey shades.}
    \label{fig:pseudo_overdensity}
\end{figure*}

In addition to the local density measured in an individual cluster, we also calculate the local density in a stacked cluster. While the local densities measured in an individual cluster reflect the actual galaxy distribution in that cluster, those measured in a stacked cluster are useful to see the relationship between the observed anisotropy in quiescent fraction and the average satellite distribution in our sample. We consider the normalised position defined as equations~\eqref{eq:newpos_x} and \eqref{eq:newpos_y}, and projected distance in units of $r_\mathrm{200m}$ instead of physical distance. Here, we apply a smoothing scale of $0.05\times r_\mathrm{200m}$ to equation~\eqref{eq:kd}. In this case, galaxies with $\delta\gtrsim 2$ are mainly distributed within $0.1\times r_\mathrm{200m}$, where satellites along the two axes might be mixed in this scale. As shown in Fig.~\ref{fig:pseudo_overdensity}, the overall trends are similar to what we see in the case of individual clusters: at a fixed overdensity over $-1<\delta<2$, the quiescent fraction along the major axis exceeds that along the minor axis except for the lowest redshift bin.

These results support the idea that the modulation in the quiescent fraction is not just a reflection of different local environments between the two axes but also a different degree of the quenching effect independent of local galaxy density. There is a possibility that the galaxy distribution does not trace that of dark matter or the ICM. Different distributions of such ``invisible'' components from the satellites possibly lead anisotropic quenching due to anisotropic strengths of tidal stripping or ram-pressure stripping. Indeed, \citet{Martin-Navarro2021} have argued that the ICM density in a halo is anisotropic due to the AGN feedback of the central galaxy.

\subsection{Anisotropic quenching versus stellar mass}
\label{sec:mass}
To discuss the physical origins of anisotropic quenching, stellar mass is another important factor. This is because quiescent fraction strongly depends on stellar mass (e.g. \citealp{Peng2010,Darvish2016}) and different environmental quenching mechanisms depend on stellar mass differently (e.g. \citealp{Balogh2016,Boselli2022}). For example, ram-pressure stripping works more efficiently for less massive galaxies due to their shallower gravitational potential well.

\begin{figure*}
	\includegraphics[width=2\columnwidth]{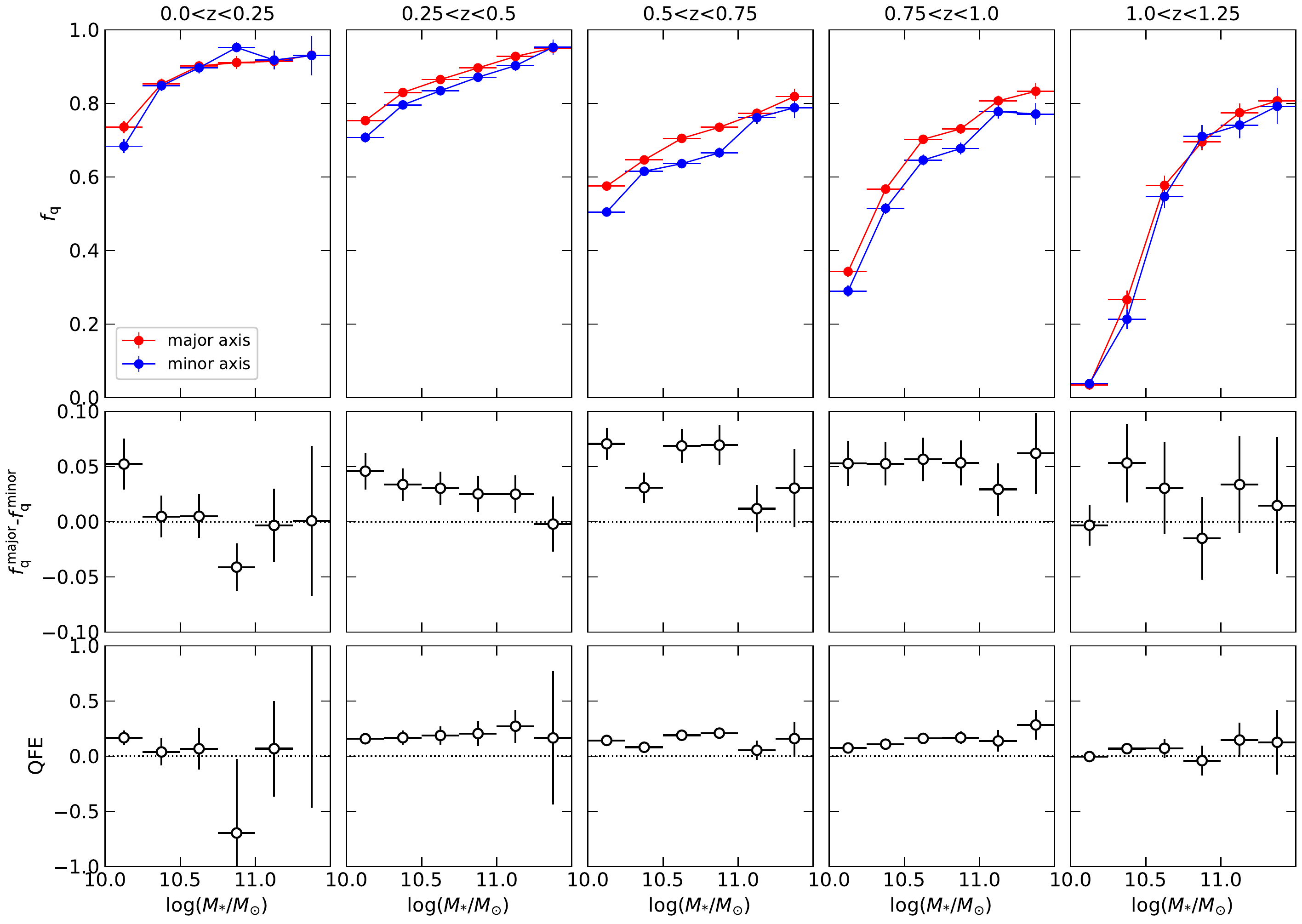}
    \caption{\textit{Top}: The quiescent fraction as a function of stellar mass. Red and blue symbols indicate the major and minor axis directions with an opening angle of $30\degree$, respectively. Each column shows different redshifts. \textit{Middle}: The Difference of quiescent fraction between the two axes at each stellar mass bin. \textit{Bottom}: The QFEs as a function of stellar mass.}
    \label{fig:mass}
\end{figure*}

To understand the stellar mass dependence of anisotropic quenching, we calculate quiescent fraction as a function of stellar mass along the two axes separately. We show the derived quiescent fractions and the difference between the two axes in the top and middle panels of Fig.~\ref{fig:mass}, respectively. The quiescent fractions along both of the two axes increase with stellar mass. At $0.25<z<1$, quiescent fractions are higher along the major axis over a wide stellar mass range. In the bottom panels of Fig.~\ref{fig:mass}, we show the quiescent fraction excess (QFE) defined as:
\begin{equation}
    \mathrm{QFE}(M_{*})=\frac{f_\mathrm{q}^\mathrm{major}(M_{*})-f_\mathrm{q}^\mathrm{minor}(M_{*})}{1-f_\mathrm{q}^\mathrm{minor}(M_{*})},
\end{equation}
where $f_\mathrm{q}^\mathrm{major/minor}(M_{*})$ is the quiescent fraction along the each axis as a function of stellar mass. This quantity means what fraction of SFGs along the minor axis would be quenched by additional environmental effects if they were located along the major axis \citep{vandenBosch2008,Peng2010,Balogh2016}. At $0.25<z<1$, QFEs are basically positive and show no clear stellar mass dependence. This implies that the environmental effects that cause the observed anisotropic quenching equally affect satellites with any stellar mass. We discuss the interpretation of it in \S~\ref{sec:origin}.

\subsection{Contributions of stellar mass and local density distributions}
As we see in \S~\ref{sec:local_density} and \S~\ref{sec:mass}, the observed anisotropy clearly depends on stellar mass and local environment: satellites with higher stellar masses or those located at denser environments have higher quiescent fractions. This means that the anisotropy in quiescent fraction may naturally occur if satellites' stellar mass and/or local density distributions are different between the two axes. Although we confirm that anisotropy still exists even at a fixed stellar mass or local density, it is still worth quantifying to what degree the observed anisotropy can be explained by the differences in stellar mass and local density.

First, we calculate the quiescent fraction of cluster galaxies within $r_\mathrm{200m}$ regardless of orientation angle as a function of stellar mass and the local density. Then we take a weighted mean of the quiescent fraction using the stellar mass and local density distributions along each axis:
\begin{flalign}
    \label{eq:exp_anisotropy_1}
    &\Delta f_{\mathrm{q}}^\mathrm{pred} =f_{\mathrm{q,major}}^\mathrm{pred}-f_{\mathrm{q,minor}}^\mathrm{pred},  \\[6pt]
    \label{eq:exp_anisotropy_2}
    &f_{\mathrm{q},k}^\mathrm{pred} =\frac{\sum_{j}\sum_{i} f_{\mathrm{q}}(M_{*i}, \delta_{j}) \cdot n_{k}(M_{*i}, \delta_{j})} {\sum_{j} \sum_{i} n_{k}(M_{*i}, \delta_{j})},
\end{flalign}
where $f_{\mathrm{q}}(M_{*i},\delta_{j})$ is the quiescent fraction at a given stellar mass and overdensity bin with intervals of $0.1\,\mathrm{dex}$ for $M_{*i}$ and $0.25\,\mathrm{dex}$ for $\delta_{j}$, respectively, and $n^{j}(M_{*i},\delta_{j})$ is the number of galaxies along the axis of $k=\{\mathrm{major}|\mathrm{minor}\}$ at $(M_{*i}, \delta_{j})$. Considering two environmental measures (i.e. overdensities for ``individual'' or ``stacked'' clusters, see \S~\ref{sec:local_density}), we calculate $\Delta f_\mathrm{q}^\mathrm{pred}$ for five cases: when only stellar mass is considered, when only environment is considered (``individual'' or ``stack''), and  when both stellar mass and environment (``individual'' or ``stack'') are considered.

\begin{figure}
	\includegraphics[width=\columnwidth]{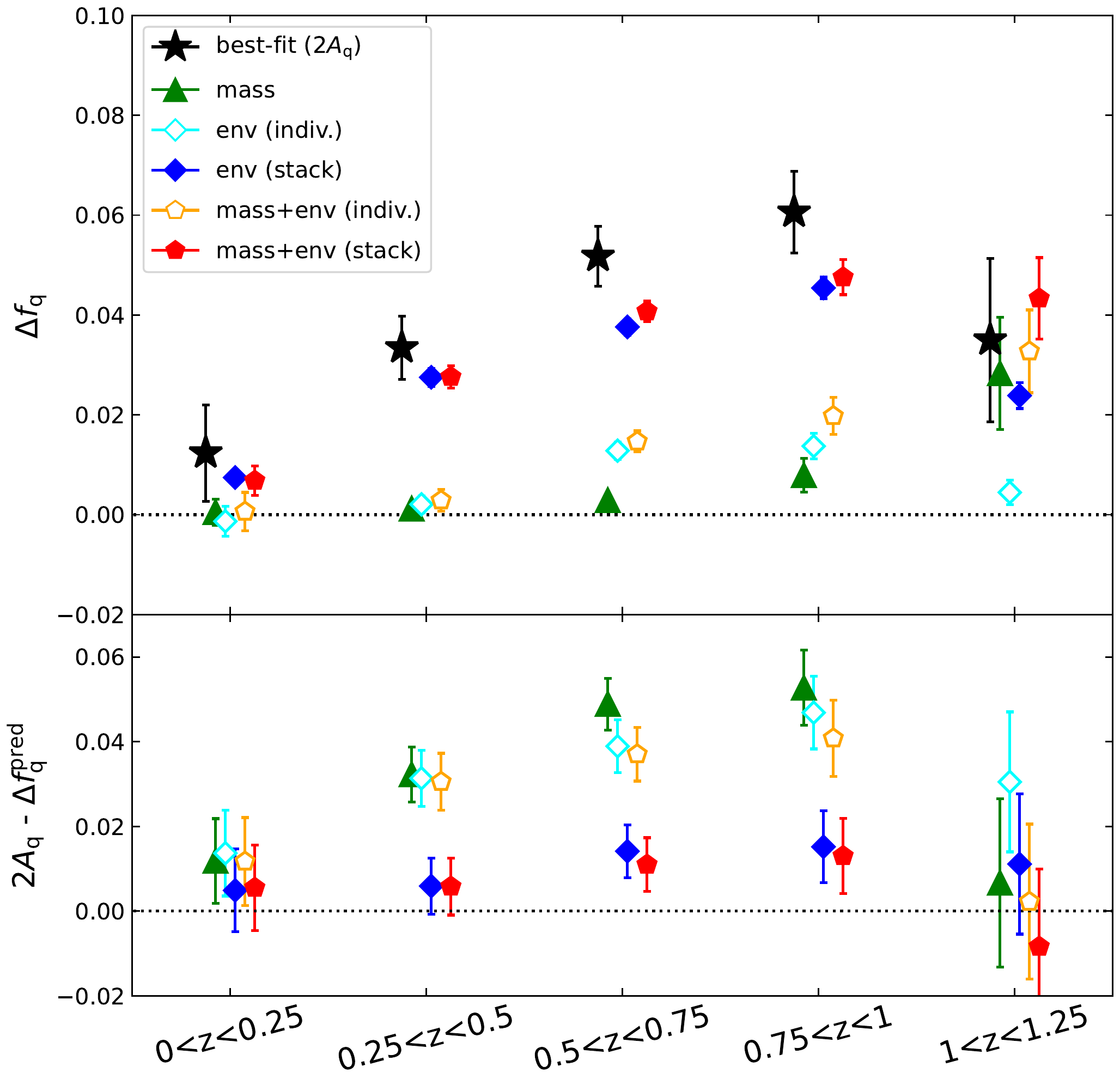}
    \caption{\textit{Top:} The expected anisotropy signals $\Delta f_{\mathrm{q}}^\mathrm{pred}$ calculated from equations~\eqref{eq:exp_anisotropy_1} and \eqref{eq:exp_anisotropy_2}. The Horizontal position corresponds to five redshift bins. $\Delta f_{\mathrm{q}}^\mathrm{pred}$ are calculated for five cases: only stellar mass differences (triangle), only the local density differences with two local density definitions (diamonds) and both stellar mass and local density differences (pentagons). Open and filled symbols indicate the local densities calculated within ``individual'' and ``stacked'' clusters, respectively. Black stars are the amplitudes of the observed anisotropy derived from the best-fit sinusoidal models (i.e. $2A_\mathrm{q}$). \textit{Bottom:} The excesses of the observed anisotropy from the predicted ones ($\Delta f_{\mathrm{q}}^\mathrm{pred}$). The meaning of each symbol is the same as the top panel.
    }
    \label{fig:intrinsic_anisotropy}
\end{figure}

We show $\Delta f_\mathrm{q}^\mathrm{pred}$ in the top panel of Fig.~\ref{fig:intrinsic_anisotropy}. Black stars are the amplitudes of the observed anisotropy derived from the best-fit sinusoidal models (i.e. $2A_\mathrm{q}$). At $0.25<z<1$, the contribution of stellar mass distribution to the observed anisotropy is negligible. Similarly, the contribution of the difference in the local density is small if we adopt the overdensity measured in ``individual'' clusters. However, if we consider the overdensity measured in ``stacked'' clusters, the expected anisotropy becomes larger and closer to the observed values. The combination of stellar mass and local density shows a similar trend: the predicted anisotropy is significantly larger when we consider the overdensity of the ``stacked'' case.

In the bottom panel of Fig.~\ref{fig:intrinsic_anisotropy}, we show the excesses of the observed anisotropy from those expected from differences in stellar mass and/or local density distributions. At $0.5<z<1$, the observed anisotropy clearly exceeds the expected ones, suggesting that the anisotropy in quiescent fraction with amplitudes of $\sim 1\text{--} 4$ per cent is purely due to additional quenching effects working differently along the two axes. At $0.25<z<0.5$, the observed anisotropy also exceeds the predicted ones, although the excess could be more tentative depending on the definition of local density. At $z<0.25$ and $z>1$, the excess is even more unclear. These results support that, at least $0.25<z<1$, the anisotropic quenching is not an apparent effect produced by the well-known correlation between star formation rate and stellar mass and/or environment.

\subsection{Overall spatial distributions of cluster galaxies}
\label{sec:overall}
\begin{figure*}
	\includegraphics[width=2\columnwidth]{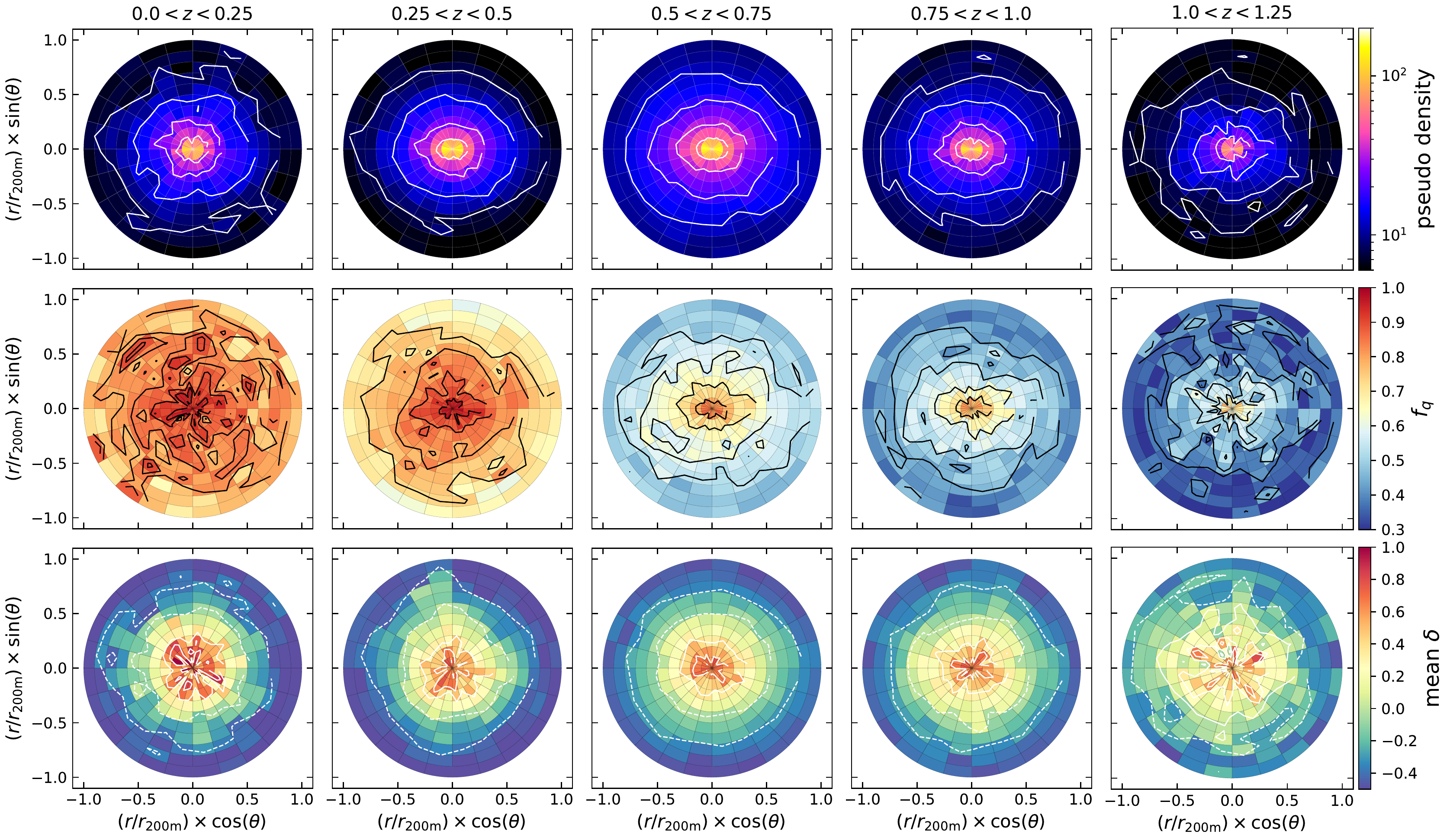}
    \caption{The spatial distribution of pseudo-density (top), quiescent fraction (middle), and mean overdensity measured in individual clusters (bottom). Each column shows different redshifts. The grid is drawn in polar coordinates as $\Delta r/r_\mathrm{200} =0.1$ and $\Delta \theta =15\degree$. Contours in each panel indicate $90,\ 75,\ 50$, and $25$ percentiles.}
    \label{fig:polar_prof}
\end{figure*}

In this paper, we examine the relationships between quiescent fraction and orientation angle, cluster-centric-radius, and local density. We visually summarise them in Fig.~\ref{fig:polar_prof}. We divide the stacked cluster into $240$ spatial bins with $24$ angular bins and $10$ radial bins. In each bin, we calculate the quiescent fraction and the average of local overdensity measured in individual clusters. To see the link between the direction of the major axis of central galaxies and the average cluster morphology, we also calculate a ``pseudo-density'' map with a unit of $r_\mathrm{200m}^{-2}$ by dividing the number of galaxies per cluster in each bin by the area. We also plot contours showing $90,\ 75,\ 50$, and $25$ percentiles in each panel. We note that field subtraction has not been applied.

For every redshift bin except for $0<z<0.25$, the distribution of the quiescent fractions is elongated along the horizontal (major axis) direction especially at small radii, which is observed as anisotropic quenching. This trend has been reported by \citet{Martin-Navarro2021} at $z<0.08$. The pseudo-density is also elongated in the same way as the quiescent fraction, suggesting that the cluster morphology is aligned to the direction of the central's major axis. This is consistent with many studies which have reported an alignment between the major axis of central galaxies and the cluster shape (e.g. \citealp{Yang2006,Huang2016}). In contrast to the overall galaxy density profiles of pseudo-clusters, the average local overdensity measured in individual clusters are somewhat symmetric rather than elongated.

\subsection{Origins of anisotropic quenching}
\label{sec:origin}
In \S~\ref{sec:result}, we show that anisotropic quenching clearly exists at least at $0.25<z<1$. \citet{Martin-Navarro2021} have proposed that the weakened ram-pressure stripping due to diluted ICM densities along the minor axis by the AGN feedback of centrals is the origin of anisotropic quenching. On the other hand, \citet{Stott2022} has pointed out the possibility that anisotropy in quiescent fraction can be observed without the AGN feedback if the cluster morphology is elongated towards the major axis of the central galaxy, because environmental quenching depending on matter density works more strongly at a fixed cluster-centric-radius in that direction. Considering the discussion above, we attempt to constrain on the physical origins of anisotropy.

We first discuss the possibility that elongated cluster morphology causes anisotropic quenching. From the local density dependence discussed in \S~\ref{sec:local_density}, we find that anisotropy exists even if the local galaxy density is fixed. This suggests that anisotropic quenching may not be just due to the difference in the underlying matter density. However, even if the local density at the time of observation is fixed, it is possible that the average local densities through which satellites pass during accretion might be different between the two axes. Galaxies infalling from the direction of the cluster's elongation pass through higher density environments before arriving at a given local density than those from the perpendicular direction. Therefore, satellites along the major axis of cluster morphology may experience environmental effects more strongly during accretion and be more likely to be quenched. To test this scenario, we need to investigate the relationship between the local density during accretion, cluster ellipticity and orientation angle, which might be done with a cosmological simulation.

To discuss the physical origins of anisotropic quenching other than cluster morphology, we review what we find in this study. As described in \S~\ref{sec:radial}, anisotropic quenching is mainly observed within $r_\mathrm{200m}$, suggesting that the physical mechanisms causing anisotropy work efficiently within cluster haloes rather than outside of clusters as pre-proccesing. This is supported by \citet{Martin-Navarro2021}, who have shown that the majority of galaxies contributing to the anisotropy are those quenched after infalling into cluster haloes. 

In \S~\ref{sec:mass}, we see stellar mass dependence of anisotropy. \citet{Martin-Navarro2021} have reported that lower-mass galaxies show a larger modulation in quiescent fraction than higher-mass galaxies: $0.027\pm 0.002$ at $\log(M_{*}/M_{\odot})<10.5$ and $0.014\pm 0.002$ at $\log(M_{*}/M_{\odot})>10.5$, which can be converted to peak-to-peak differences of $\sim 0.054$ and $\sim 0.028$ for the low-mass and high-mass bins. In Fig~\ref{fig:mass}, there is a sign of a similar trend at $0.25<z<0.5$: $0.041 \pm 0.011$ at $10<\log(M_{*}/M_{\odot})<10.5$ and $0.027\pm 0.010$ at $10.5<\log(M_{*}/M_{\odot})<11.5$. In this sense, lower-mass satellites have a larger contribution to the anisotropic signal at least in the lower-redshift sample. However, such a trend is not seen at $z>0.5$. This might be due to a redshift evolution or just due to uncertainties in stellar mass estimates.

Another important point is that the difference in the QFEs between the two axes has little or no stellar mass dependence, implying that the physical mechanisms causing anisotropic quenching equally affect satellites regardless of their stellar masses. This seems to conflict with the scenario which attributes ram-pressure stripping as the primary origin because ram-pressure stripping preferentially quenches less massive galaxies, which have shallower gravitational potential wells (e.g. \citealp{Abadi1999,Yan-Tan2022,Boselli2022}). This might suggest that the observed anisotropic quenching is caused by a combination of multiple quenching effects rather than by a single dominant mechanism. Another possibility is that the stellar mass limit adopted in this study ($10^{10}\,M_{\odot}$) is too high to see such mass dependence. \citet{Yan-Tan2022} have found that the QFE for disc-like galaxies in massive clusters at $0.3<z<0.6$ decreases with stellar mass at $8.5<\log(M_{*}/M_{\odot})<10.5$ and reported such QGs with a disc-like morphology are thought to be quenched through ram-pressure stripping. If we focus on such low-mass satellites, we might also detect a decreasing trend of the QFE with stellar mass.

Lastly, we discuss the physical origins of anisotropic quenching in terms of typical timescales of quenching mechanisms. As a necessary condition for anisotropic quenching to be observed, the quenching mechanism causing it must have a shorter timescale than the time it takes for the satellites to change their angular position against their centrals' major axis. As an upper limit, we assume that satellites keep their orientation during only one cluster crossing time \citep{Kuchner2022}, roughly estimated to be $\sim 2\,\mathrm{Gyr}$ \citep{Boselli2006}. The crossing time may be even shorter at higher redshifts due to smaller cluster sizes.  The typical timescale of ram-pressure stripping is $\lesssim 0.5\text{--}1.0\,\mathrm{Gyr}$ \citep{Boselli2022}, satisfying the condition. Other quenching mechanisms representative in clusters have longer timescales than the crossing time: $\sim 3\,\mathrm{Gyr}$ for harassment (e.g. \citealp{Moore1998}) and $\gtrsim7\,\mathrm{Gyr}$ for strangulation (e.g. \citealp{Boselli2006,Boselli2014d}). Therefore, only anisotropic quenching due to ram-pressure stripping can be observed under the assumption that satellites stay in the same orientation against their centrals during one infalling or back-splashing time.

Although this simple comparison prefers ram-pressure stripping as the primary origin of anisotropic quenching, there are some uncertainties to be considered. The quenching timescale depends on properties of clusters (e.g. halo mass, size and ICM density) as well as those of satellites (e.g. gas mass, size, morphology). If satellites have large orbital angular momenta, the upper limit of the timescale that anisotropy is visible would be shorten. Moreover, galaxy discs with the edge-on inclination to the direction of motion are by $\sim 50$ per cent less effectively stripped of their gas by ram pressure than those with the face-on inclination \citep{Abadi1999}, leading to a longer quenching timescale. Anisotropy in quiescent fraction naturally occurs if the quenching timescale of a certain quenching mechanism depends on satellites' orientation against the central galaxies. To discuss these points, tracing the trajectories of satellites in a cosmological simulation will be valuable.

\section{Summary}
We have investigated the anisotropic quenching of cluster satellite galaxies over $0<z<1.25$ using more than $5000$ clusters from the CAMIRA cluster catalogue and a galaxy catalogue with photo-\textit{z} and stellar-mass estimates from the HSC-SSP. Thanks to deep photometry of the HSC-SSP, galaxies are almost complete down to $1\times 10^{10}\,M_{\odot}$. We have calculated quiescent satellite galaxy fraction as a function of the orientation angle measured from the major axis of central galaxies. The main results are as follows.

\begin{enumerate}
  \item At $0.25<z<1$, the quiescent fraction shows clear fluctuations with several per cent amplitudes, which is periodically higher and lower in the major ($\theta\sim0\degree,\,180\degree,\,360\degree$) and minor ($\theta\sim90\degree,\,270\degree$) axis directions, respectively. We fit the quiescent fraction data with sinusoidal functions and detect non-zero amplitudes of the modulation at $>5\sigma$ level, demonstrating the presence of anisotropic quenching. We have confirmed that sinusoids fit the data much better ($\chi_{\nu}^{2}\sim1$) than constants ($\chi_{\nu}^{2}\gtrsim2$). At $z<0.25$ and $z>1$, we find no clear evidence of anisotropic quenching possibly due to the small sample sizes.
  
  \item Satellite galaxies are clearly aligned towards the major axis of their centrals regardless of star formation category with a slightly weak concentration of SFGs than QGs. These different degrees of angular segregation cause the observed anisotropy in quiescent fraction.
  
  \item The radial profiles of quiescent fraction around both the major and minor axes decrease toward the cluster outskirt except for the lowest redshift bin. While the quiescent fractions are higher for the major axis direction within $r_\mathrm{200m}$, those around the two axes are almost the same in the cluster outskirt ($>r_\mathrm{200m}$). This trend implies that the physical mechanisms causing anisotropic quenching mainly work within the cluster halo rather than within larger-scale filaments outside of the cluster as pre-processing. 

  \item Even when the local galaxy density is fixed, the quiescent fraction along the major axis at $0.25<z<1$ is slightly but significantly higher than that along the minor axis. This suggests that the observed anisotropic quenching cannot be fully explained by the difference in the local density between the two axes.

  \item The differences in the QFE between the two axes are almost independent of stellar mass. A difference in the strength of ram-pressure stripping between the two axes, which is one of the candidates of physical origins of the observed anisotropy, is not consistent with this mass-independence because ram-pressure stripping works more efficiently for lower-mass galaxies. There is a possibility that our stellar mass limit in this study is too high to see such mass dependence clearly. It can also be possible that anisotropic quenching is caused by multiple environmental effects, making mass dependence invisible.
  
  \item The observed anisotropy clearly exceeds the predicted anisotropy caused by the differences in stellar mass and/or local density distributions between the two axes. This supports that the anisotropic quenching is not fully explained by a stellar mass segregation or a skewed satellite distribution towards the centrals' major axis direction and needs additional environmental effects which work differently along the two axes.
 
  \item For a signature of anisotropic quenching to be observed, satellites must be quenched on a timescale shorter than the time over which their initial orientation angles against their centrals' major axis are significantly changed. We have assumed this timescale to be one cluster crossing time, $\sim 2\,\mathrm{Gyr}$. While the typical quenching timescale of ram-pressure stripping is less than $1\,\mathrm{Gyr}$, those of harassment and strangulation are $\sim 3\,\mathrm{Gyr}$ and $\gtrsim7\,\mathrm{Gyr}$, respectively. Therefore, among these three representative environmental effects in clusters, ram-pressure stripping is preferred.
\end{enumerate}

We have not detected a clear signature of anisotropic quenching at $z<0.25$ and $z>1$ nor in the cluster outskirt at any redshift. These no-detection results may be due to the intrinsic weakness of anisotropy but also due to large uncertainties in field subtraction. Spectroscopic confirmation of member galaxies may increase the possibility of the detection of anisotropy.

In addition, it is important to estimate how long satellite galaxies can actually maintain their initial orientation angles. We propose tracing the trajectories and orientations of satellites in a cosmological simulation. This will help us examine whether ram pressure-stripping is a likely mechanism of anisotropic quenching as expected and whether harassment, strangulation and other mechanisms should also be included as candidates.

\section*{Acknowledgements}
We appreciate the anonymous referee for constructive comments and suggestions. We also thank Suin Matsui and Takumi Tanaka for valuable comments and discussions at meetings. MA acknowledges support from Japan Science and Technology Agency (JST) Support for Pioneering Research Initiated by the Next Generation (SPRING), Grant Number JPMJSP2108 and Japan Society for the Promotion of Science (JSPS) KAKENHI Grant Number 22J11975. KS acknowledges support from JSPS KAKENHI Grant Number JP19K03924.

The Hyper Suprime-Cam (HSC) collaboration includes the astronomical communities of Japan and Taiwan, and Princeton University. The HSC instrumentation and software were developed by the National Astronomical Observatory of Japan (NAOJ), the Kavli Institute for the Physics and Mathematics of the Universe (Kavli IPMU), the University of Tokyo, the High Energy Accelerator Research Organisation (KEK), the Academia Sinica Institute for Astronomy and Astrophysics in Taiwan (ASIAA), and Princeton University. Funding was contributed by the FIRST program from the Japanese Cabinet Office, the Ministry of Education, Culture, Sports, Science and Technology (MEXT), JSPS, JST, the Toray Science Foundation, NAOJ, Kavli IPMU, KEK, ASIAA, and Princeton University.

This paper is based on data collected at the Subaru Telescope and retrieved from the HSC data archive system, which is operated by Subaru Telescope and Astronomy Data Centre (ADC) at NAOJ. Data analysis was in part carried out with the cooperation of Centre for Computational Astrophysics (CfCA) at NAOJ.  We are honoured and grateful for the opportunity of observing the Universe from Maunakea, which has the cultural, historical and natural significance in Hawaii.

This paper makes use of software developed for Vera C. Rubin Observatory. We thank the Rubin Observatory for making their code available as free software at \url{http://pipelines.lsst.io/}. 

The Pan-STARRS1 Surveys (PS1) and the PS1 public science archive have been made possible through contributions by the Institute for Astronomy, the University of Hawaii, the Pan-STARRS Project Office, the Max Planck Society and its participating institutes, the Max Planck Institute for Astronomy, Heidelberg, and the Max Planck Institute for Extraterrestrial Physics, Garching, The Johns Hopkins University, Durham University, the University of Edinburgh, the Queen’s University Belfast, the Harvard-Smithsonian Centre for Astrophysics, the Las Cumbres Observatory Global Telescope Network Incorporated, the National Central University of Taiwan, the Space Telescope Science Institute, the National Aeronautics and Space Administration under grant No. NNX08AR22G issued through the Planetary Science Division of the NASA Science Mission Directorate, the National Science Foundation grant No. AST-1238877, the University of Maryland, Eotvos Lorand University (ELTE), the Los Alamos National Laboratory, and the Gordon and Betty Moore Foundation.

We use the following open source software packages for our analysis: \texttt{numpy} \citep{numpy:2011}, \texttt{pandas} \citep{pandas:2010}, \texttt{scipy} \citep{scipy:2001}, \texttt{astropy} \citep{astropy:2013,astropy:2018} and \texttt{matplotlib} \citep{matplotlib:2007}.

\section*{Data Availability}
The HSC-SSP PDR3 is available in the data release site at \url{https://hsc-release.mtk.nao.ac.jp/doc/}.

%%%%%%%%%%%%%%%%%%%%%%%%%%%%%%%%%%%%%%%%%%%%%%%%%%



\bibliographystyle{mnras}
\bibliography{aiq_camira_2} % if your bibtex file is called example.bib

%%%%%%%%%%%%%%%%%%%%%%%%%%%%%%%%%%%%%%%%%%%%%%%%%%

%%%%%%%%%%%%%%%%% APPENDICES %%%%%%%%%%%%%%%%%%%%%

%\appendix
%\section{Appendix A}
%\label{app:a}
%Appendix, Appendix, Appendix.

%%%%%%%%%%%%%%%%%%%%%%%%%%%%%%%%%%%%%%%%%%%%%%%%%%

% Don't change these lines
\bsp	% typesetting comment
\label{lastpage}
\end{document}